\documentclass[aps,prl,twocolumn,noshowpacs,10pt]{revtex4-1}

\usepackage{graphicx}
\usepackage{amsmath}

\begin{document}

\title{Can high-density human collective motion be forecasted \\ by spatiotemporal fluctuations?}

\author{Arianna Bottinelli}
\email{arianna.bottinelli@su.se}
\affiliation{NORDITA, Stockholm University, Roslagstullsbacken 23, SE-106 91, Stockholm, Sweden}

\author{Jesse L. Silverberg}
\email[]{js@mss.science}
\affiliation{Multiscale Systems, Inc., Division for Advanced Sciences and Data Research, Worcester, MA 01609}

\maketitle

\textbf{Concerts, protests, and sporting events are occurring with increasing frequency and magnitude.  The extreme physical conditions common to these events are known to cause injuries and loss-of-life due to the emergence of collective motion such as crowd crush, turbulence, and density waves~\cite{helbing2007dynamics, krausz2012loveparade, fruin1993causes, heide2004common}.  Mathematical models of human crowds aimed at enhancing crowd safety by understanding these phenomena are developed with input from a variety of disciplines~\cite{Helbing2000, moussaid2011simple, duives2013state}.  However, model validation is challenged by a lack of high-quality empirical data~\cite{helbing2007dynamics, berrou2007calibration} and ethical constraints~\cite{moussaid2016crowd} surrounding human crowd research.  Consequently, generalized model-based approach for real-time monitoring/risk-assessment of crowd collective motion remains an open problem~\cite{johansson2008crowd,kumar2018intelligent}.  Here, we take a model-free approach to crowd analysis and show that emergent collective motion can be forecasted directly from video data.  We use mode analysis methods from material science and concepts from non-equilibrium physics to study footage of a human crowd at an {\it Oasis} rock concert.  We analyze the attendees’ positional fluctuations during a period of crowd turbulence to predict the spatial patterns of an emergent human density wave.  In addition to predicting spatial patterns of collective motion, we also identify and measure temporal patterns that precede the density wave and forecast its appearance by $\approx 1$~s.  Looking ahead, widening this forecasting window beyond 1~s will enable new computer vision technologies for real-time risk-assessment of emergent human collective motion.
}

Crowd disasters reporting injury or loss of life are low-probability high-impact ``black swan'' events~\cite{heide2004common, helbing2007dynamics, krausz2012loveparade, moussaid2011simple, taleb2007black, fruin1993causes}.  As such, it is both rare and difficult to obtain research-quality video footage to examine how these disasters occur~\cite{helbing2007dynamics, moussaid2018virtual}.  The scarcity of empirical data~\cite{berrou2007calibration}, technical constraints of tracking algorithms~\cite{rodriguez2017analysis}, and ethical concerns surrounding experimental crowd research have thus limited study of high-density crowds where the risk and need are greatest.  Rock concerts, however, present an opportunity to address these challenges and observe crowds in extreme social conditions routinely occurring at these events~\cite{Silverberg2013, krausz2012loveparade, earl2004influences}.  By studying rock concerts, we can tackle essential questions such as what types of emergent phenomena can be found in high-density human crowds, and whether there are measurable quantities forecasting the occurrence of dangerous collective motion.  

\begin{figure*}
\centering{\scalebox{.95}{\includegraphics{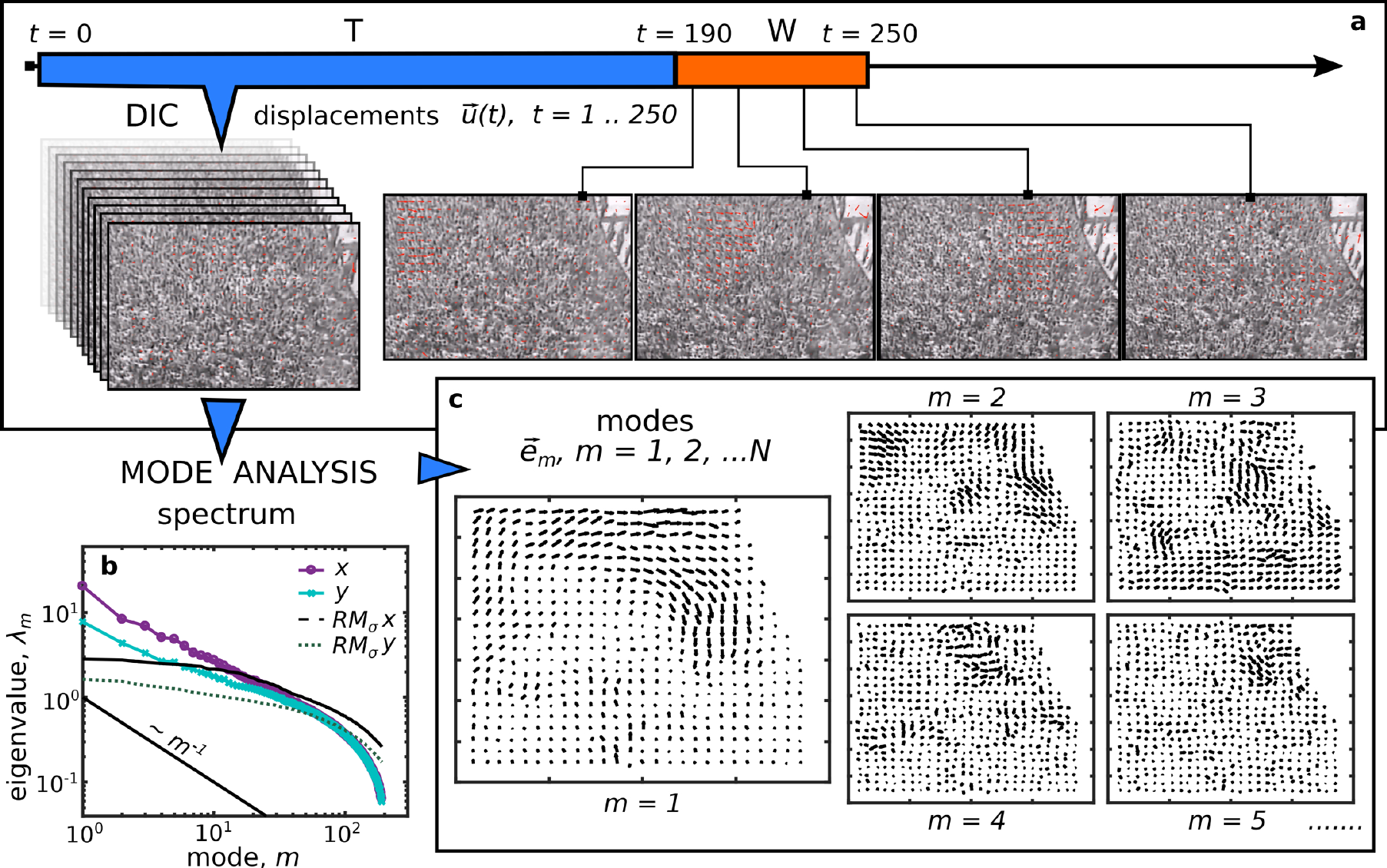}}
\caption{Mode analysis applied to crowd footage of an \textit{Oasis} concert. {\bf a}, Crowd displacements $\vec{u}(t)$ are tracked on a grid using DIC.  We divide video frames into two segments distinguished by the crowd's motion.  The segment with turbulent crowd motion $T$ includes frames $t = 1, \ldots, 190$, and the segment with a collective human density wave $W$ includes frames $t = 191, \ldots, 250$.  Black and white images show individual frames from the movie; superimposed red arrow vector fields show measured crowd displacements $\vec{u}(t)$ at each example time point.  {\bf b}, Mode analysis applied to the displacements from $T$ finds 15 modes along the $x$-direction (purple) and 73 modes along the $y$-direction (teal) that are above the noise threshold defined by the random matrix model $RM_{\sigma}$ (black, $x$- and $y$-component are dashed and dotted respectively). {\bf c}, Spatial plots of the first 5 modes ($x$- and $y$-direction combined).  Notice the $m = 1$ vector field closely matches the trajectory of the wave in $W$.}
\label{fig1}}
\end{figure*}

Here, we analyze a publicly available YouTube video of a crowd filmed by a security camera during an \textit{Oasis} concert with $\approx$~60,000 attendees in Manchester, UK (Fig.~\ref{fig1}a; Supplementary Movie 1; SI).  The footage has a 3/4 top-down perspective and shows the attendees exhibiting a fluctuating turbulent motion~\cite{helbing2007dynamics, moussaid2011simple, yu2007modeling, helbing2007crowd}, as well as two particularly large density waves that travel across the entire field-of-view, pushing attendees toward the stage~\cite{ivancevic2012turbulence}.  Upon reaching the front of the audience, these waves partially reflect off security barriers and gradually dissipate.  The camera maintains a fairly constant viewing angle and zoom for the first of these waves, offering $\approx$~42~s (353~frames at 8.33~frames/s; Methods; SI) of analyzable footage including the moments leading up to the density wave, the density wave itself, and the moments immediately after.  

We therefore segment the movie into two distinct periods corresponding to frames before the wave, $T$ (Fig.~\ref{fig1}a, blue), and during the wave, $W$ (Fig.~\ref{fig1}a, orange).  $T$ consists of the movie's first 190 frames and is characterized by turbulent crowd motion with individuals fluctuating in apparently random directions.  Toward the end of $T$, a density wave forms at a location outside the camera's field-of-view.  $W$ consists of the next 60 frames and is characterized by the wave (i) appearing at the left-most edge of the video, (ii) propagating rightward to the front of the crowd, (iii) impacting the stage's security barrier, and (iv) partially reflecting while dissipating (Fig.~\ref{fig1}a, sequence of four frames during $W$).  $W$ concludes at frame 250 with the crowd reentering a period of turbulent collective motion.  Our aim is to analyze footage recorded during $T$ (frames $t = 1 \ldots 190$) to determine what measurable quantities forecast spatiotemporal patterns of the density wave during $W$ (frames $t = 191 \ldots 250$).

Using quantitative image analysis, we measured the crowd's coarse-grained time-dependent displacement field $\vec{u}(t)$ from the concert video (Fig.~\ref{fig1}a, red vector fields; Methods; Supplementary Movie 2) \cite{Silverberg2013}.  Coarse-graining was achieved by quantifying displacements on an interpolated square grid of $N = 556$ points with digital image correlation (DIC).  The inter-grid point spacing defined square areas covering approximately 3 to 4 people  (Methods; SI).  From the displacements $\vec{u}(t)$ measured during $T$, we computed the fluctuation covariance matrix, $C_{ij} = \langle [ \vec{u}_i(t) - \langle \vec{u}_i \rangle ] \cdot [ \vec{u}_j(t) - \langle \vec{u}_j \rangle ] \rangle$, where subscripts index each grid point and $\langle \cdots \rangle$ denotes temporal averaging over the 190 frames of $T$ \cite{henkes2012extracting, bottinelli2016emergent, bottinelli2017}.  In large, dense, disordered systems such as vibrated granular media, the eigenvalues $\lambda_m$ (Fig.~\ref{fig1}b) and eigenvectors $\vec{e}_m$ (Fig.~\ref{fig1}c) of $C_{ij}$ convey information about the system's collective response to perturbations~\cite{chen2011measurement, manning2011vibrational}.  For simplicity, these eigenvectors are often referred to as the system's \textit{modes}.  Modes with eigenvalues larger than the $RM_{\sigma}$ noise threshold (Fig.~\ref{fig1}b; Methods) are vector fields exhibiting long-range coherent spatial displacements (e.g., Fig.~\ref{fig1}c, $m = 1$)~\cite{henkes2012extracting}.  At equilibrium, these modes are the most easily excitable responses to perturbations, oscillating at frequencies $\omega_m = 1/\sqrt{\lambda_m}$~\cite{ashcroft1976nd}.  In the context of dense human crowds, mode analysis is applicable at the scale of individual people, and modes can be interpreted as vector fields describing the most likely direction of collective motion \cite{bottinelli2016emergent}.  Here, due to low video definition and intrinsic challenges of high-density crowd tracking~\cite{rodriguez2017analysis}, we instead apply mode analysis at the scale of the coarse-grained grid (Methods).  This simplification is possible because large-scale collective motions are robust with respect to modest coarse graining (SI).  Thus, we turn raw video data recorded during $T$ into a crowd-specific forecast for emergent collective motion.

Within the framework of mode analysis, the first mode $\vec{e}_1$ is the most easily excitable, and therefore describes the most likely collective motion to occur~\cite{bottinelli2016emergent,xu2010anharmonic}.  For the \textit{Oasis} crowd footage, $\vec{e}_1$ has attendees in the upper quarter of the camera's field-of-view moving toward the stage, while attendees near center-stage move parallel and away from to the security barrier (Fig.~\ref{fig1}c, $m = 1$).  Remarkably, this vector field accurately matches the trajectory and reflection of the propagating density wave in $W$, appearing to function as a waveguide for the emergent collective motion.  Of critical importance, we reiterate $\vec{e}_1$ was computed from data acquired during $T$.  Therefore the first mode forecasts the spatial trajectory of the density wave before it occurs.  

\begin{figure*}
\centering{\scalebox{.95}{\includegraphics{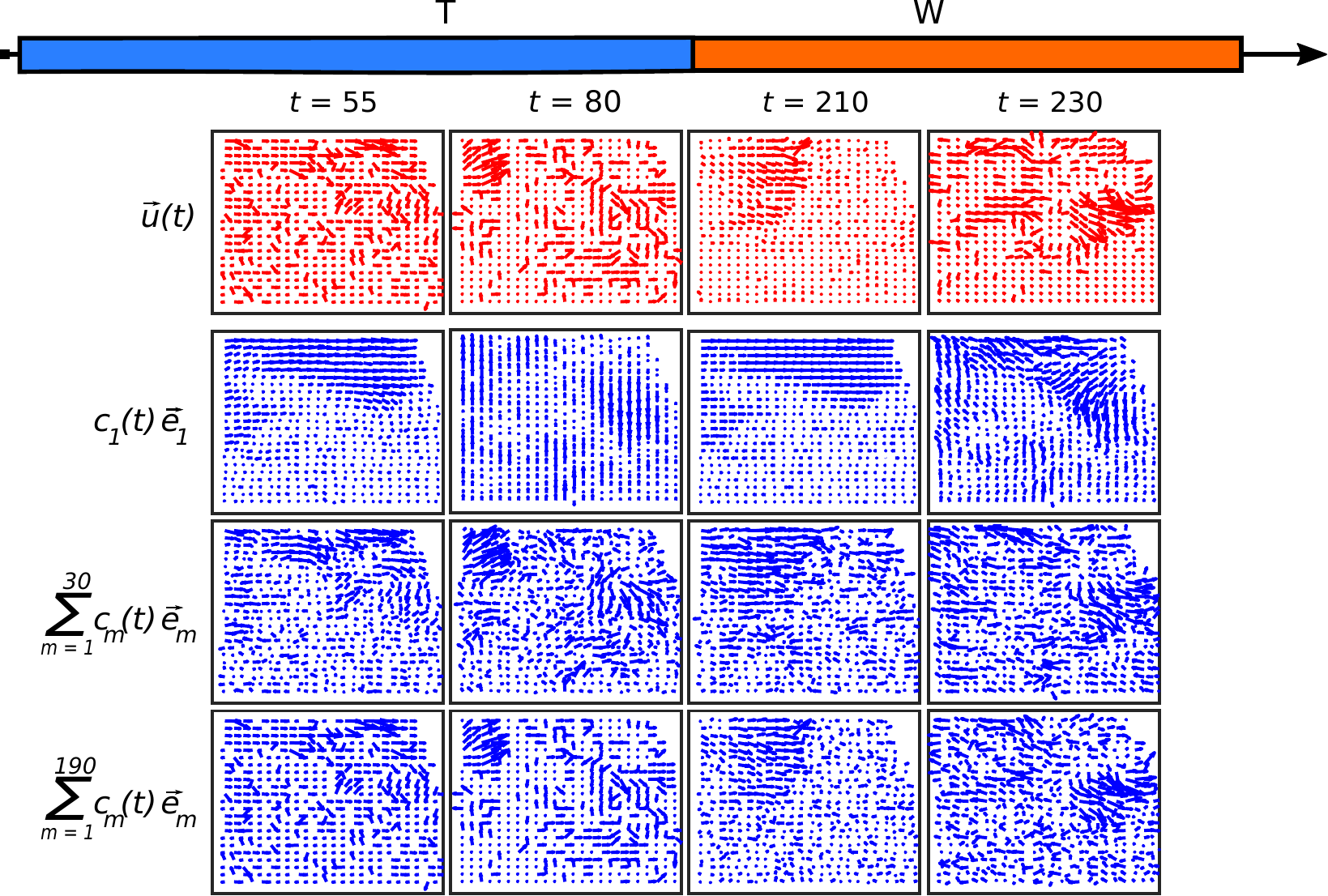}}
\caption{Spatiotemporal reconstructions (blue) of the crowd's measured displacements (red) using a subset of the computed modes.  The first mode bears little resemblance with the crowd's motion for frames during $T$, but shows better agreement during $W$.  The combination of the first 30 modes generates substantial agreement with the original displacement field during $T$ and $W$, while combining 190 modes offers only minor improvements.  Summing over 30 modes was chosen as a representative example to show the greater significance of low-$m$ modes compared to high-$m$ modes when reconstructing $\vec{u}(t)$.}
\label{fig2}}
\end{figure*}

In addition to explaining spatial patterns of collective motion, we use the modes to decompose the crowd's displacements and study the temporal dynamics leading to the density wave.  This decomposition is possible because the modes $\vec{e}_m$ form an orthogonal basis such that $\vec{u}(t) = \sum_{m} c_m(t) \vec{e}_m$, where the coefficients are $c_m(t) = \vec{u}(t) \cdot \vec{e}_m$, and each mode's $x$- and $y$-components are separately normalized to 1.  A partial reconstruction of $\vec{u}(t)$ summing $m$ up to 1, 30, and 190 modes illustrates the relative significance of low-$m$ modes compared to high-$m$ modes (Fig.~\ref{fig2}; SI).  For example, snap-shots of the temporal reconstruction before and during the wave show a substantial difference between a total of 1 and 30 modes, but subtle differences between 30 and 190 modes.  Thus, the combination of the first 30 modes offers a greater amount of spatial information than the combination of the next 160 modes (SI).  

\begin{figure*}
\centering{\scalebox{.95}{\includegraphics{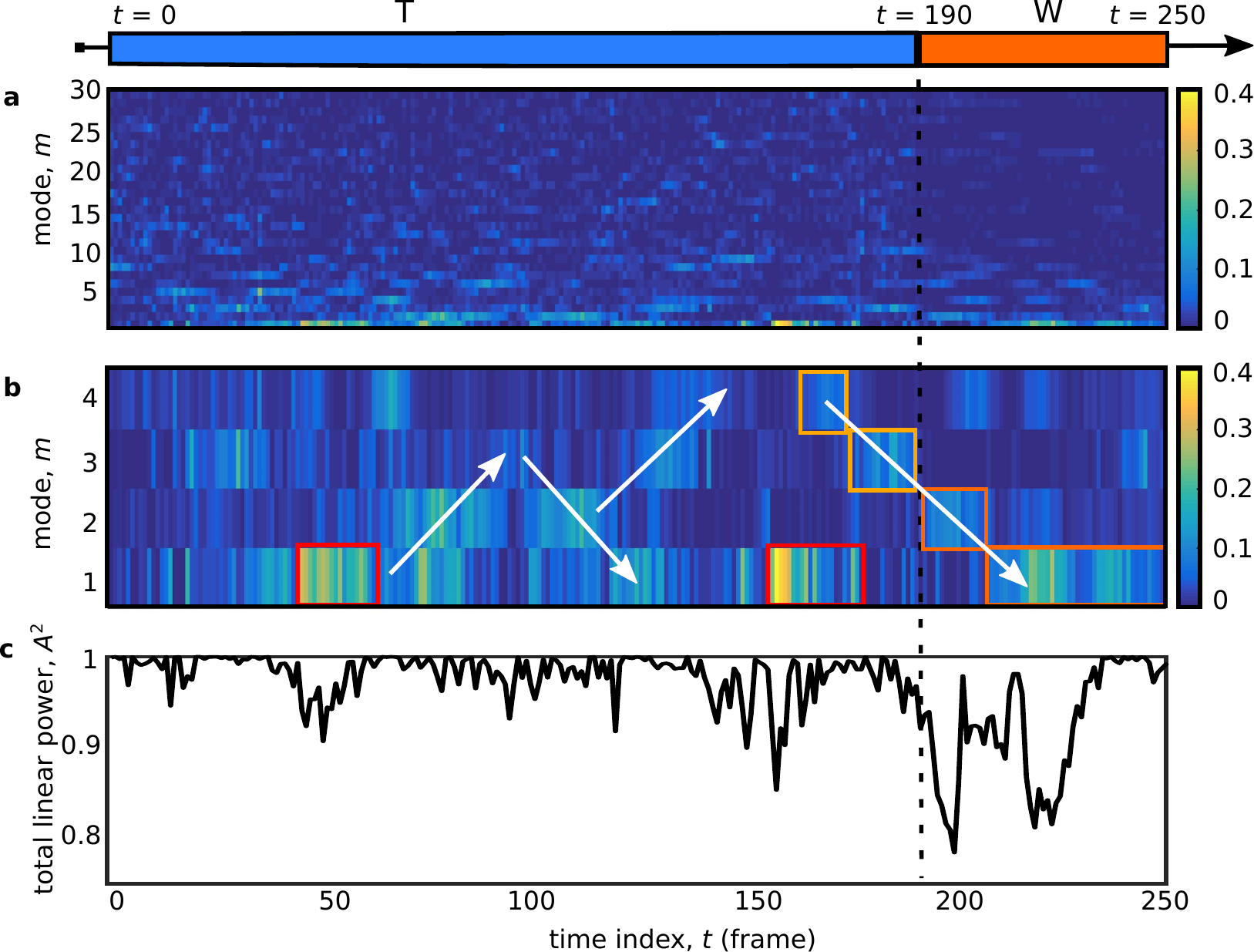}}
\caption{Time traces of the mode's explanatory power suggests spatiotemporal fluctuations can be used to forecast collective motion.  {\bf a}, The explanatory power $\alpha^2_m(t)$ (larger values are lighter; smaller values are darker) shows turbulent crowd motion during $T$ is a superposition of modes, while human density wave during $W$ is dominated by the $m = 1$ mode.  {\bf b}, Zooming in on the time-series data for modes $m = 1, \ldots, 4$ shows coherent temporal patterns of explanatory power reminiscent of cascades (arrows) that coincide with power injections (e.g., at $t \approx 50$ and $150$, outlined red).  An inverse-cascade starting during $T$ and extending into $W$ (outlined orange) becomes detectable approximately 1~s before the wave is in the camera's field-of-view, suggesting a window of opportunity to forecast the wave before it occurs.  {\bf c}, The time series for the total explanatory power of linear mode analysis $A^2(t)$ is notably smaller than 1 during moments that coincide with  power injection and collective motion, indicating non-linear mode-mode coupling plays a role during these events.}
\label{fig3}}
\end{figure*}

To quantify the amount of information conveyed by each mode, we compute the explanatory power $\alpha^2_m(t) = c_m^2(t) / 2 \left| \vec{u}(t) \right|^2 $~\cite{chen2011measurement, schreck2011repulsive}.  This quantity takes values between 0 and 1, and measures the fraction of the displacement vector field $\vec{u}(t)$ accounted for by each mode $\vec{e}_m$.  Time-traces of the explanatory power show three features of note.  (i) At the transition from $T$ to $W$, $\alpha_m^2(t)$ decreases in magnitude for modes $m > 15$ (Fig.~\ref{fig3}a).  This change in the distribution of explanatory power is a consequence of computing modes with data up to $t = 190$, and therefore reflects the difference between past and future displacements when projecting on $\vec{e}_m$ (SI).  (ii) The second notable feature is a concentration of power on the waveguide-like mode $\vec{e}_1$ during $W$, reflecting the occurrence of the density wave (Fig.~\ref{fig3}b, highlighted orange box).  Surprisingly, we discover similarly strong concentrations at $t \approx 50$ and $150$ that don't correspond to any large-scale wave in the crowd (Fig.~\ref{fig3}b, highlighted red boxes).  Rewatching the video, we instead find these two bursts of the first mode's explanatory power correspond to a group of $\approx 20$ attendees in the upper-middle region of the camera's field-of-view displacing along the direction of $\vec{e}_1$ (Supplementary Movie 2).  (iii)  The third notable feature occurs after $t \approx 50$, where $\alpha^2_m(t)$ time-traces exhibit a pattern reminiscent of direct-cascades (low-$m$ to high-$m$) and inverse-cascades (high-$m$ to low-$m$) in turbulent flows (Fig.~\ref{fig3}b, arrows)~\cite{korotkevich2008simultaneous}.  Interestingly, a similar phenomenon has been observed in vibrated granular media when large power injections on a single mode trigger power transfer between several modes due to non-linear mode-mode coupling \cite{schreck2011repulsive}.  Exploiting the analogy between human crowds and granular media~\cite{helbing2006analytical, cristiani2011multiscale, faure2015crowd}, we interpret the localized displacements along $\vec{e}_1$ at $t \approx 50$ and 150 as injections of power due to attendees pushing toward the stage, and subsequent cascades as transfer of power between low-$m$ modes.  

To account for power injections and cascades, we expand the displacement field beyond linear order to $\vec{u}(t) = \sum_m c_m(t) \vec{e}_m + \sum_{jk\ell} c_{jk\ell}(t) (\vec{e}_j \otimes \vec{e}_k) \vec{e}_{\ell} + \vec{\chi}(t)$.  The first term is a familiar linear expansion, the second term expresses mode-mode coupling as a tensor product with coefficients $c_{jk\ell}(t)$, and the third term represents power injections and dissipation due to crowd activity.  Thus, the total linear power $A^2(t) = \sum_m \alpha^2_m(t)$ ranges between 0 and 1, and quantifies the proportion of motion explained by traditional linear mode decomposition (Fig.~\ref{fig3}c).  When $A^2 < 1$ non-linear mode coupling and the crowd's activity are relevant for explaining crowd dynamics.  For the \textit{Oasis} concert crowd studied here, the explanatory power of linear mode analysis during $T$ has two notable dips to 0.9 at $t \approx 50$ and 150.  The timing of these dips coincides with power injections on $\vec{e}_1$, and the subsequent power cascades among low-$m$ are signaled by smaller deviations around 0.95.  We conclude time traces of $\alpha^2_m(t)$ are useful for detecting spatiotemporal patterns of collective motion that cannot be described by linear mode analysis.  In particular, the second power injection at $t \approx 150$ transitions into a smooth inverse-cascade from $m = 4$ that starts before the density wave enters the camera's field-of-view and ends in a concentrated signal on $m = 1$ during $W$ (Fig.~\ref{fig3}b, arrow crossing dashed line between $T$ and $W$).  Based on the available data, this inverse-cascade retrospectively offers $\approx 1$~s advance warning before the wave appears (SI). 

We therefore ask: can high-density human collective motion be forecasted by spatiotemporal fluctuations?  The analysis presented here indicates the answer is \textit{yes}, but for a relatively short window of opportunity.  Analyzing additional datasets will enable deep statistical characterization of the frequency, duration, intensity, and speed of power cascades and injections.  Correlating these statistical measures with the likelihood of emergent collective motion has the potential to transform our retrospective analysis into a real-time forecasting computer vision technology.  Such advances could reduce rates of injury and loss of life by enabling rational interventions targeted at preventing crowd disasters at mass gatherings.

\section{Movie Captions}
\begin{itemize}
  \item Supplementary Movie 1 - Raw camera footage from \textit{Oasis} concert security camera.  
  \item Supplementary Movie 2 - Analyzed camera footage from \textit{Oasis} concert security camera showing measured displacement field overlaid in red. Video encoded at 8.0 fps.
\end{itemize}

\section{Acknowledgments}
\begin{acknowledgments}
Simulations were performed on resources provided by the Swedish National Infrastructure for Computing (SNIC) at Lunarc.  AB was supported by the NORDITA Fellowship grant.  JLS was independently supported.
\end{acknowledgments}

\section{Author Contributions}
AB and JLS equally contributed to designing the research, performing the analysis, interpreting the results, and writing the manuscript.

\section{Competing interests}
The authors have no conflicts of interest to report.

\vfill \eject

\clearpage

\section{Methods}

\subsection{Digital Image Correlation (DIC)}

Video footage of the \textit{Oasis} concert crowd was downloaded and saved in  .mp4 file format.  The crowd's motion recorded in this video was quantified using previously published DIC techniques developed to study high-density human collective behavior \cite{Silverberg2013}.  In short, this method quantifies motion by comparing consecutive pairs of movie frames to measure the local displacement field $\vec{u}(t)$ on a discrete grid of $\langle x, y \rangle$ points.  

The downloaded video was encoded at 25 frames per second (fps), making it difficult to detect motion at this frame rate and zoomed-out view.  In particular, displacements of attendees are typically sub-pixel in magnitude at the $1/25$~s time scale.  The video was therefore down-sampled 3-fold from its original recording to 8.33 fps by discarding 2 out of every 3 frames, enabling the image tracking algorithm to more precisely detect motion while losing the minimal number of frames.

Motion between consecutive down-sampled frames was tracked on a coarse-grained grid.  The distance between each grid point was $\Delta = 14$ pixels in both $x$- and $y$-directions.  DIC was computed by tracking the average motion within square tiles (edge length $\Delta$) centered on each grid point.  Values for $\Delta$ from 10 to 15 pixels were tested and found to have no substantial effect on downstream analysis (SI).  

Occasional bright flashing lights recorded in the video reduce tracking efficiency due to sudden changes in pixel value intensity.  To address this difficulty, the overall brightness of each color channel was normalized before DIC was performed.  

Perspective distortions due to the camera's 3/4 top-down positioning were corrected for with a nonaffine transformation so that the typical size of attendees was constant throughout the entire field-of-view.  Specifically, the characteristic size of a human head was $\approx 7$ pixels.  

Subtle horizontal camera motion during filming induced a drift of the entire field-of-view.  We therefore utilized stationary objects in the stage's security zone (Fig.~\ref{fig1}a, top-right corner) as reference points to subtract time-dependent spatially homogeneous displacements.  As a side-effect of this camera motion, grid points near the edge of the field-of-view were not visible throughout the entirety of the movie; they were therefore discarded from analysis.  

After completing all of these data preparation steps, the video stream ingested by the DIC algorithm sampled motion on 556 grid points uniformly spread on a 315 pixels $\times$ 380 pixels region (Fig.~\ref{fig1}, Supplementary Movie 2).

\subsection{Mode analysis}
The general procedure to perform mode analysis of high-density crowds is described elsewhere~\cite{bottinelli2016emergent, bottinelli2017}.  The only methodological difference is here we apply mode analysis to the displacements $\vec{u}(t)$ measured with DIC on the coarse grained grid rather than to the displacements of individuals within the crowd. 

\subsection{Data Availability}
All data and software will be made freely available upon request.

\vfill \eject

\clearpage

%%%%%%%%%% Merge with Supplementary materials %%%%%%%%%%
\pagebreak
\begin{widetext}
\begin{center}
\textbf{\large Supplementary Information: Can high-density human collective motion be forecasted by spatiotemporal fluctuations?}\\
\end{center}
\begin{center}
\end{center}
\end{widetext}
%%%%%%%%%% Merge with Supplementary materials %%%%%%%%%%
%%%%%%%%%% Prefix a "S" to all equations, figures, tables and reset the counter %%%%%%%%%%
\setcounter{equation}{0}
\setcounter{figure}{0}
\setcounter{table}{0}
\setcounter{page}{1}
\makeatletter
\renewcommand{\theequation}{S\arabic{equation}}
\renewcommand{\figurename}{SUPPLEMENTARY FIG.}
\renewcommand{\thefigure}{\arabic{figure}}
\renewcommand{\bibnumfmt}[1]{[#1]}
\renewcommand{\citenumfont}[1]{#1}
%%%%%%%%%% Prefix a "S" to all equations, figures, tables and reset the counter %%%%%%%%%%

\twocolumngrid

\section{Description of video data}
\textit{Oasis} concert video footage was accessed from YouTube.com (\verb!https://youtu.be/BgpdmAtbhbE! ; Supplementary Movie 1).  It was recorded on July 2$^{\rm nd}$, 2005, at Manchester Stadium in Manchester, UK.  This concert venue has a capacity for 60,000 people, making it among the largest venues in the UK.  The official audio-video recording is publicly available on YouTube.com (\verb!https://youtu.be/y5zZ-yHM1MU!), and it shows footage of both the band and the crowd from several perspectives, offering a well-rounded impression for the context of the event. 

Footage used for analysis was recorded by two security cameras, M8 C36 and M4 C34, which show overhead views of the concert's audience between roughly 21:00:00 and 21:06:00.  The first camera, M8 C36, is placed to the right side of the stage and shows in sequence: turbulent motion, a density wave (Supplementary Movie 1, playback time 1:22, corresponding to camera time-stamp 21:05:27), more turbulent motion, and a second density wave (Supplementary Movie 1, playback time 2:20, corresponding to camera time-stamp 21:06:25).  Both density waves start from the back of the crowd outside the camera's field of view, move toward the stage, impact the security barriers separating the audience from the band, and subsequently dissipate.  Part of the wave is reflected by the barriers while part propagates along them.  The second camera, M4 C34, is pointed at the same region of the crowd as the first camera, but from the left side of the stage.  It initially shows the crowd turbulence occurring before the first wave (Supplementary Movie 1, playback time 3:14; note the time gap from 21:01:49 to 21:04:36), the first wave (Supplementary Movie 1, playback time 4:06), and concludes by zooming out while recording wave propagation. 

For our analysis, we selected a 42~s sequence from camera M8 C36, which includes footage of crowd turbulence and the first wave (Supplementary Movie 1, playback time 0:55 to 1:37).  During this sequence, the camera maintains a constant zoom, but exhibits transient and gradual longitudinal drift due to camera panning.  This global translational motion is corrected for by subtracting the apparent motion of stationary objects in the camera's field of view (Methods).

\section{Displacement-based measurements} 

Statistical properties of the crowd's displacement vector field $\vec{u}(t)$ have unambiguous signatures that distinguish turbulent collective motion from a human density wave.  Here, we demonstrate this finding with a simple measure of the displacement field's \textit{mean and standard deviation}, as well as a more sensitive measure using the \textit{displacement fluctuation correlation length}.

\begin{figure*}
\centering
\includegraphics{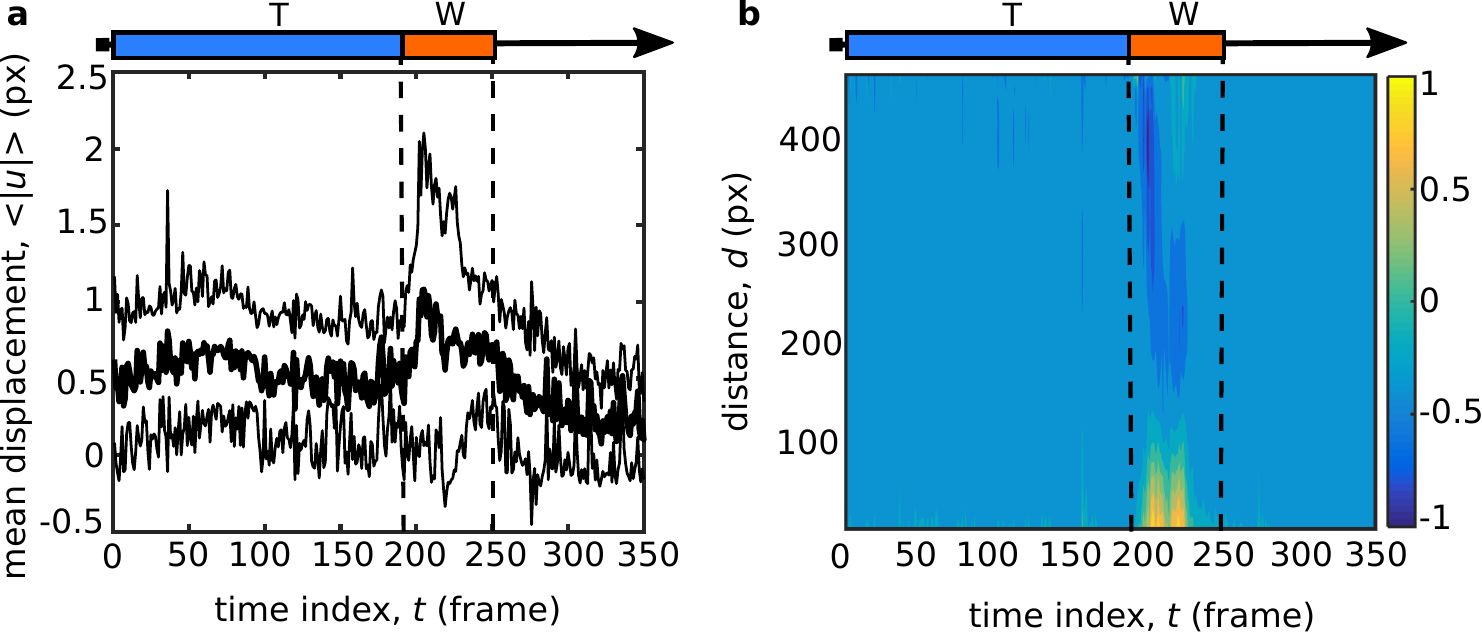}
\caption{\small Spatial measures on the displacement vector field $\vec{u}$. {\bf a}, Mean (thick line) plus and minus the standard deviation (thin lines) of the absolute displacement as a function of time. {\bf b}, Correlation function of the fluctuations around the mean absolute displacement as a function of distance and time.}
\label{fig:spatial}
\end{figure*}

For each time sample where we measured the crowd's displacement vector field $\vec{u}(t)$, we computed the displacement mean $\bar{u}(t) = \big\langle \left| \vec{u}(t) \right|_i \big\rangle$ and standard deviation ${\rm STD}\big( \left| \vec{u}(t) \right|_i \big)$ (Supplementary Fig.~\ref{fig:spatial}a).  These statistical properties are computed on the lengths of individual displacement vectors sampled at each of the $i = 1, \dots, 556$ grid points.  Taking the mean as an example, we clarify that this calculation is the \textit{average of each vectorial displacement's norm}, not the \textit{norm of the average vectorial displacements}.  We found that displacements during crowd turbulence typically fluctuate around $(0.5 \pm 0.5)$~pixels (Supplementary Fig.~\ref{fig:spatial}a, data from time index during $T$), which is notably smaller than the distance $\Delta$ between DIC grid points (Methods, $\Delta = 14$~pixels), and similarly smaller than the typical human head size ($\approx 7$~pixels).  In contrast, the average displacements during the density wave are larger and more heterogeneous (Supplementary Fig.~\ref{fig:spatial}a, data from time index during $W$).  

The spatial correlation function of the displacement's fluctuations about their mean value is defined as
\begin{equation}
C(d,t) = \Big\langle \ \big[ \vert \vec{u}_i(t) \vert -  \bar{u}(t) \big] \times \big[\vert \vec{u}_j(t)\vert -  \bar{u}(t) \big] \ \Big\rangle_{d_{ij}=d}, \nonumber
\end{equation}
where the average $\langle \cdots \rangle_{d_{ij}=d}$ is over all pairs of grid points $i$ and $j$ a distance $d$ apart.  $C(d,t)$ takes values between -1 and 1 and quantifies the instantaneous spatial coherence of displacements.  We find $C(d,t)$ during crowd turbulence decays over a very short distance and becomes slightly negative, indicating displacements are random and not significantly coherent (Supplementary Fig.~\ref{fig:spatial}b, data from time index during $T$).  In contrast, $C(d,t)$ during the density wave has strong correlations up to $\approx 120$~pixels and long-range anti-correlation, demonstrating the density wave's collective and coherent nature (Supplementary Fig.~\ref{fig:spatial}b, data from time index during $W$).  Thus, like the mean and standard deviation of displacements, the spatial correlation function shows a clear statistical signature distinguishing two types of crowd behavior with a sharp transition between them. 

We further note that while $C(d,t)$ shows a subtle peak at $t \approx 150$, none of the displacement-based measures explored in this section clearly detect the group of $\approx 20$ attendees collectively moving toward the stage at $t \approx 50$ and $150$.  This missed-detection is because globally averaged quantities are insensitive to localized crowd motions that occur in a small portion of the field of view.

\section{Intermediate mode analysis results}

\begin{figure*}
\centering
\includegraphics{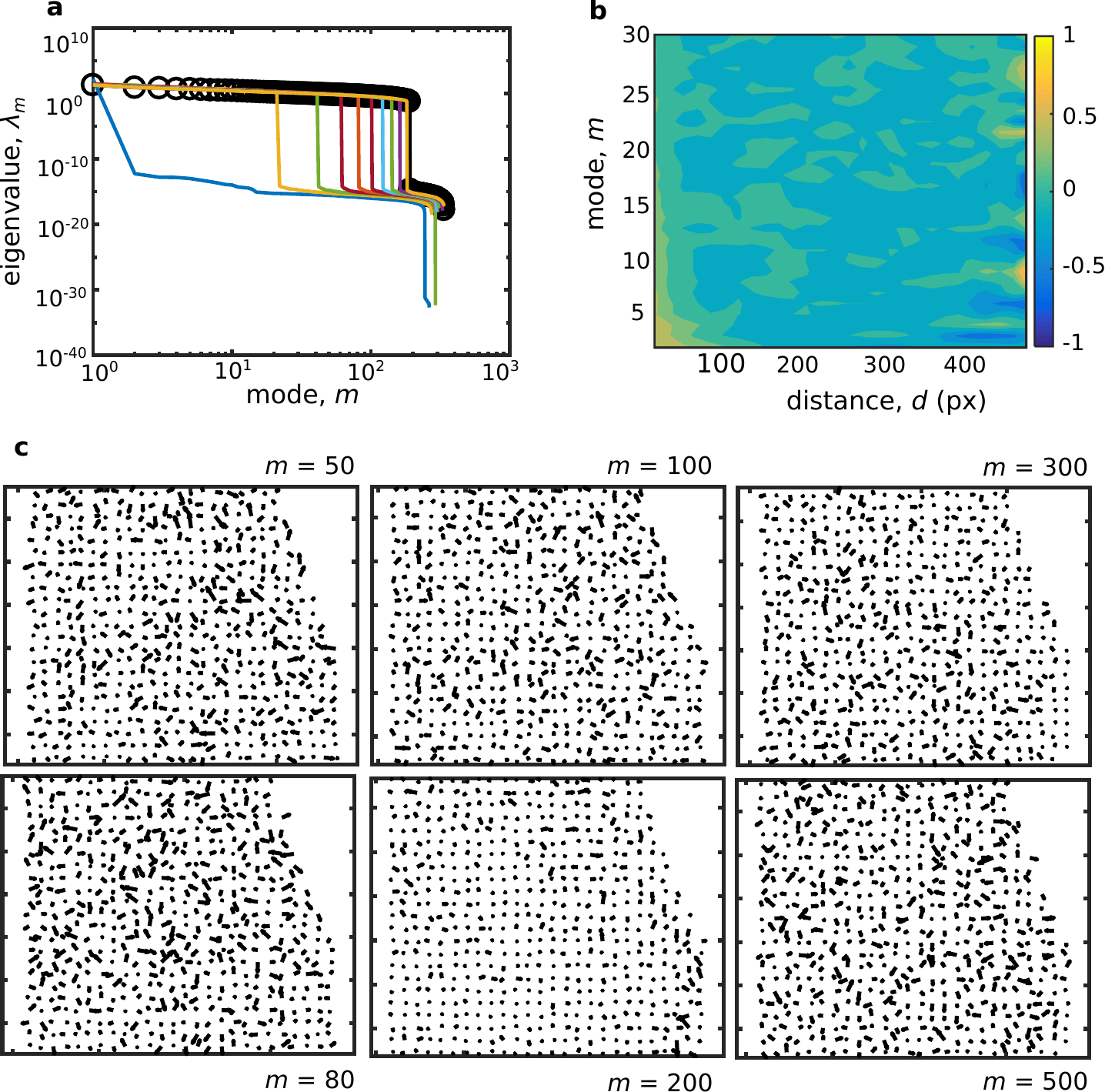}
\caption{\small Additional results of mode analysis. {\bf a}, Dependence of the eigenvalue spectrum on the number of frames used for computing $C_{ij}$.  Circles represent the spectrum computed by using all available $t_T = 190$ frames.  The first blue line uses only one frame for computing $C_{ij}$, each following line corresponds to 20 additional frames ($t_T = 1, 21, 41, \ldots$).  Only the $x$-component of the spectrum is shown for clarity.  {\bf b}, Spatial coherence $C_m(d)$ for the first 30 modes quantified by the correlation function for mode's fluctuations about its mean orientation. {\bf c}, Eigenmodes at high values of $m$ are qualitatively similar.}
\label{fig:modeanalysis}
\end{figure*}

Conventionally, mode analysis is applied to trajectories of individuals within crowds to study the group's collective behavior~\cite{bottinelli2016emergent, bottinelli2017}.  This analysis is a multi-step method where intermediate calculations are important for validation, even though they are not essential for understanding the final results.  In this work, a key innovation that we made was to apply mode analysis to the crowd's displacements measured on a coarse-grained grid, thus overcoming the challenge of acquiring individual trajectories (Methods).  Intermediate results of mode analysis presented here therefore have additional significance beyond validation of the method, because they offer evidence and support for its extension to a new context.  

In order to adapt conventional mode analysis for this work, we treated each point $i$ in the coarse-grained grid as if it was an individual in conventional mode analysis.  We then computed the displacement covariance matrix $C_{ij}$ averaged over the $t_T = 190$ frames of $T$, where subscripts index each of the $N = 556$ grid points.  This matrix's eigenvalues $\lambda_m$ and eigenvectors (modes) $\vec{e}_m$, where $m = 1, \dots, 2N$, provide the essential information about the crowd's collective response to perturbations.  

Modes corresponding to collective motion are determined by comparing the eigenvalue spectrum of $C_{ij}$ to the eigenvalue spectrum of the corresponding random matrix model $RM_{\sigma}$~\cite{bottinelli2016emergent, bottinelli2017}.  Of the 556 modes computed along the $x$-direction, 15 were larger than the $RM_{\sigma x}$ noise threshold.  Likewise, of the 556 modes computed along the $y$-direction, 73 were larger than the $RM_{\sigma y}$ noise threshold.  Thus, $\approx 8\%$ of all modes computed from the coarse-grained grid data correspond to collective motion, similar to the $3\%$ found when using mode analysis to study high-density crowd simulations wherein full trajectory data were available~\cite{bottinelli2016emergent}.  

Of the total $2N = 1112$ modes obtained from mode analysis, 380 had eigenvalues larger than $10^{-2}$ (190 along the $x$-direction, $190$ along the $y$-direction), 341 eigenvalues were between $10^{-10}$ and 0, and the remaining 391 were negative (211 along $x$ and 180 along $y$, all of similar magnitude).  The presence of discontinuities in the eigenvalue spectrum reflects the fact we do not have enough time frames during $T$ for the spectrum of $C_{ij}$ to fully converge to the correlation matrix $C_p$~\cite{henkes2012extracting}.  

Convergence of $C_{ij}$ can be quantitatively checked by verifying the inequality $r = 2N / t_T < 1.5$, which in the present case gives $r = 2 \times 556 / 190 = 5.9$.  Even though the $r < 1.5$ criteria was unmet, we found we were still able to extract the low-$m$ modes corresponding to long-range coherent collective motion.  We determined this result by computing the eigenvalue spectrum of $C_{ij}$ while varying the number of frames used in its computation (Supplementary Fig.~\ref{fig:modeanalysis}a, compare black circles for all $t_T = 190$ frames, to colored lines using data subsets with less than 190 frames).  Through this analysis we find (i) the eigenvalues corresponding to low-$m$ modes are stable with respect to removal of time frames, and (ii) the absence of convergence mainly affects high-$m$ modes that happen to be below the $RM_{\sigma}$ threshold separating coherent from random collective motion.  Thus, the predictions for collective crowd motion, which rely on the lowest-$m$ modes (Fig.~3b, up to $m = 4$), are robust with respect to the limited amount of available data.  

A typical intermediate step when performing mode analysis is to identify and discard rattlers~\cite{bottinelli2016emergent,bottinelli2017}.  These are isolated members within the aggregate caged by their surrounding neighbors.  Being relatively free to move, rattlers do not participate in collective motion and their presence leads to readily identifiable large eigenvalues.  Here, the coarse-grained grid spatially averages over the motion within a square region of $\Delta^2$ pixels.  Thus, we do not expect to detect rattlers in the traditional sense of under-constrained individuals.  However, there is a generalized notion of rattlers wherein a DIC grid point may detect uncorrelated motion for a variety of reasons such as (i) a cluster of under-constrained people, (ii) a tracking error, or (iii) the presence of a physical barrier segregating a region of the crowd.  It is therefore reasonable to check for rattlers in this general sense, even though the specific underlying mechanism may not have the conventional interpretation.  In our case the eigenvalue spectrum does not present any anomalously large eigenvalues characteristic of rattlers (Supplementary Fig.~\ref{fig:modeanalysis}a, black circles, spectrum varies smoothly at low-$m$).

We characterized the spatial coherence of each mode by computing the correlation function for the displacement's fluctuations around their mean orientation,
\begin{equation}
C_m(d) = \Big\langle \ \big[ \vec{e}_m^{\,i} - \vec{\Phi}_m \big] \cdot \big[ \vec{e}_m^{\,j} - \vec{\Phi}_m \big] \ \Big\rangle_{d_{ij}=d}, \nonumber
\end{equation}
where $\vec{\Phi}_m = N^{-1} \sum_{i=1}^N \vec{e}_m^{\,i}/ \vert \vec{e}_m^{\,i} \vert $ is the mean direction of the $m^{\rm th}$ mode.  Examining $C_m(d)$ shows that low-$m$ modes are long-range correlated, and spatial coherence decreases with mode number (Supplementary Fig.~\ref{fig:modeanalysis}b).  We further note that higher-$m$ modes are qualitatively similar and show no long-range spatial coherence, regardless of the size or sign of their corresponding eigenvalue (Supplementary Fig.~\ref{fig:modeanalysis}).

\section{Stability with respect to the choice of grid spacing}

Motion was tracked on a coarse-grained square grid with spacing equal to $\Delta$.  While we presented results in the main text for $\Delta = 14$ pixels, we also tested values between 10 and 15 pixels.  This range of values was determined by balancing three competing constraints.  

First, the mode analysis convergence criteria $r = 2 N / t_T < 1.5$, indicates that for the available number of time frames $t_T$ the number of grid points should be $N \le 142$.  For the analyzed 315 pixel $\times$ 380 pixel region, this criteria roughly translates to a coarse-grained grid with rows and columns spaced by $\Delta_{C1} \approx 28$~pixels.  

Second, DIC imposes an optimality condition such that the grid size should be comparable to the characteristic tracking feature size.  For the \textit{Oasis} concert footage, the characteristic features within the camera's field of view are human heads, which gives a grid spacing $\Delta_{C2} \approx 7$ pixels.  

Third, a constraint comes from tracking large-scale collective motion itself, and it sets the grid size to be roughly 1/10 or smaller the characteristic size of the density wave to achieve reasonable spatial resolution.  Given a wavelength $\approx 100$ pixels, this constraint indicates $\Delta_{C3} \le 10$ pixels.  

It's not possible to simultaneously satisfy all three constraints on $\Delta$.  Because our previous analysis showed the convergence criteria for $\Delta_{C1}$ was a soft constraint that did not affect our main results (Supplementary Fig.~\ref{fig:modeanalysis}a and related discussion), we tested the stability of our results on the intermediate range $\Delta = 10, \ldots, 15$.

\begin{figure*}
\centering
\includegraphics{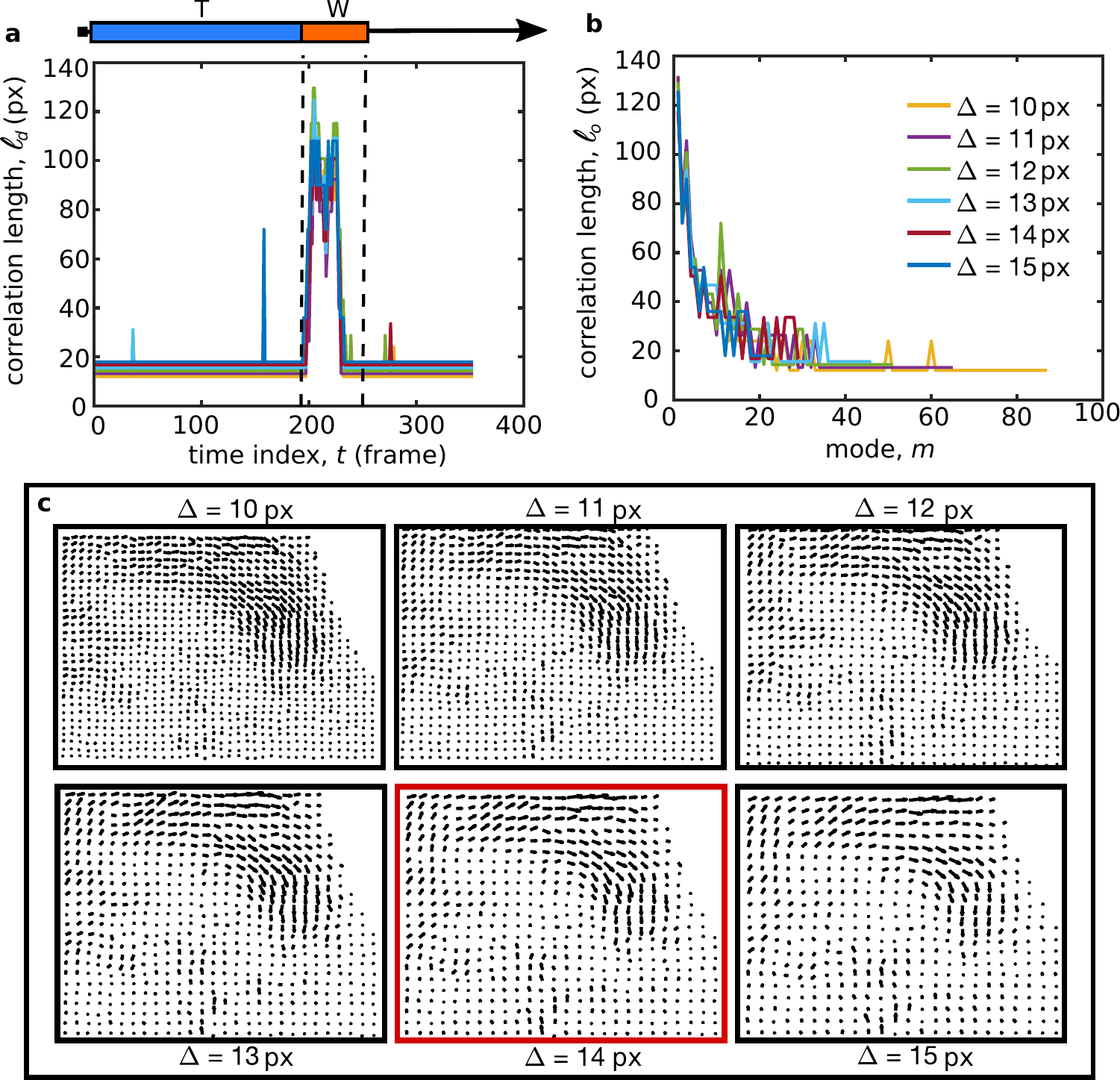}
\caption{\small Spatial properties and mode analysis do not significantly depend on the choice of the grid spacing $\Delta$. {\bf a}, Correlation length of the displacements fluctuations, {\bf b}, correlation length of the fluctuations around the mean orientation for various modes, and {\bf c}, the first mode's vector field $\vec{e}_1$ as a function of different values of $\Delta$.}
\label{fig:delta}
\end{figure*}

To quantitatively determine the influence of $\Delta$, we recall the correlation length $\ell$ of a correlation function $C(d)$ is defined as the minimum distance at which $C(\ell) = 0$.  With this metric of spatial characterizations in mind, we measured (i) the time-dependent correlation length $\ell_d$ of the displacement's fluctuations around their average length (Supplementary Fig.~\ref{fig:delta}a), and (ii) the mode-dependent correlation length $\ell_o$ of the average orientation on each mode $m$ for this range of values of $\Delta$ (Supplementary Fig.~\ref{fig:delta}b).  In both cases, we found no notable $\Delta$-dependence.  Moreover, we found the first mode $\vec{e}_1$ is qualitatively self-consistent when varying $\Delta$, even though the spatial sampling changes considerably from $\Delta = 10$ pixels to 15 pixels (Supplementary Fig.~\ref{fig:delta}c, total number of sampled points varies more than 2-fold).  

As a whole, these results indicate mode analysis on a coarse-grained grid is highly stable for grid sizes within the tested range.  We therefore selected $\Delta = 14$ as a well-rounded compromise that suitably captures small-scale details without severely overstepping the convergence criteria.

\section{Detailed mode decomposition results}

We projected the dynamics $\vec{u}(t)$, $t = 1, \dots, 250$, on the orthogonal basis of modes $\vec{e}_m$ according to $c_m(t) = \vec{u}(t) \cdot \vec{e}_m$.  Notice that this is a formal expression, and in practice, we separate the $x$- and $y$-components according to $c^x_m(t) = \sum_{i = 1}^{N} e^{\, i, x}_m u^{\, i, x}(t)$, where the sum is over all grid points $i = 1, \dots, N$, and there is an analogous expression for the $y$-component.  This component-wise separation for calculating $2N$ coefficients is necessary because modes in the $x$- and $y$-directions have unique eigenvalues and therefore cannot simply be assumed to have the same coefficient in linear mode decomposition.  This notational formality explains why in the main text, our presentation of the spatiotemporal reconstructions appears to have only one coefficient for $\vec{e}_1$, but the $x$- and $y$-directionality change sign independently (Fig.~2, first row of blue vector fields).  Moreover, each mode's $x$- and $y$-components are separately normalized such that $\left| e_m^{2} \right| = \left| \Sigma_i e_{m,i}^{2} \right| = 1$, where the sum is over the grid points $i$.  As a consequence, the value of the projection coefficients $c_m$ depends on how much of the dynamics project onto each mode, as well as the instantaneous absolute value of the displacements $\vert \vec{u}(t) \vert$.

\subsection{Reconstruction of the dynamics}

\begin{figure*}
\centering
\includegraphics{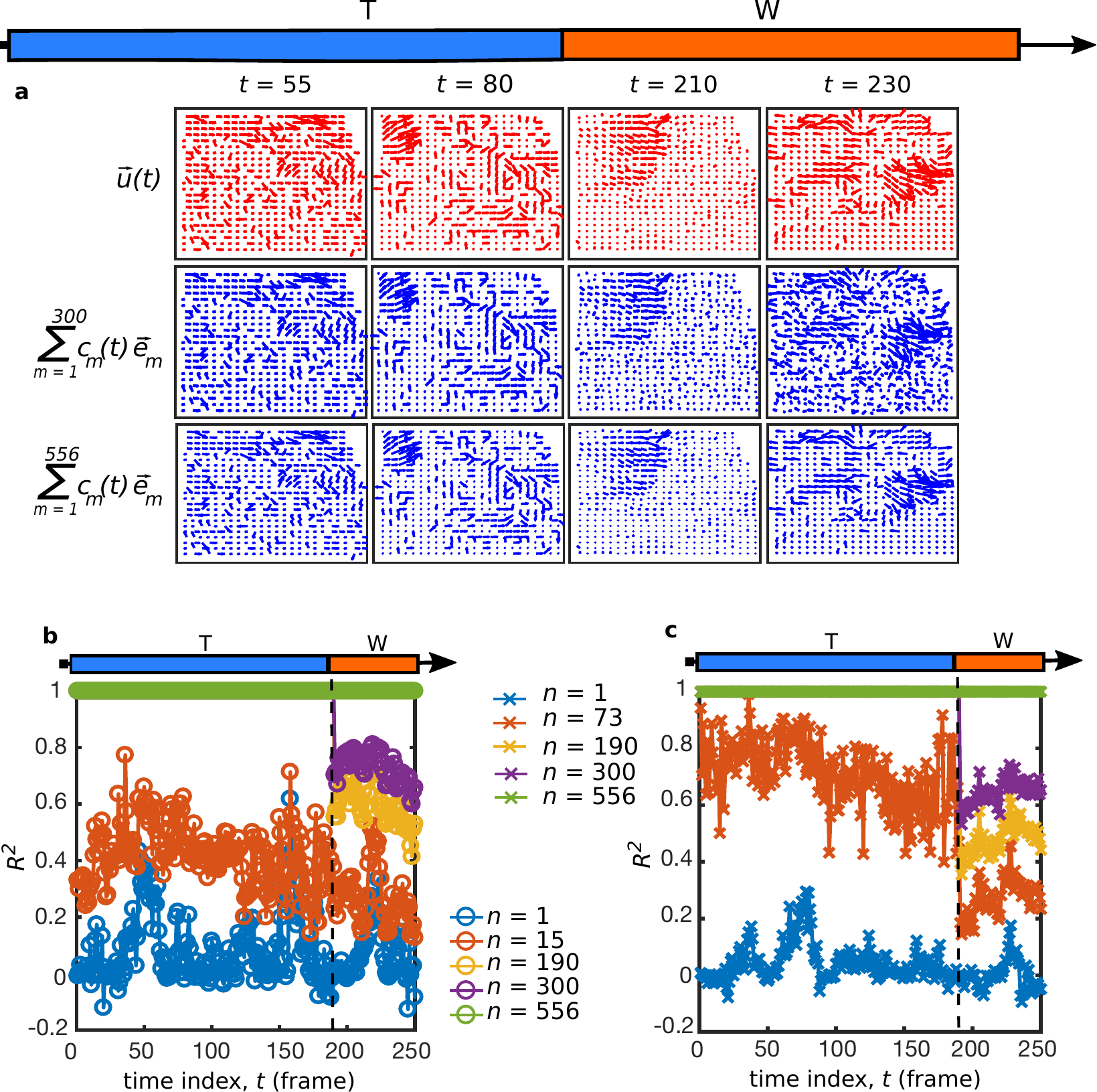}
\caption{\small Reconstruction of the dynamics with varying number of modes. {\bf a}, Comparison between the full dynamics $\vec{u}(t)$ (red vector fields) and reconstructed dynamics $\vec{u}^{(n)}(t)$ (blue vector fields) at selected time frames $t$.  Vector fields show a partial reconstruction with $n = 300$ modes, and a complete reconstruction with $n = 556$ modes. The time-dependent statistic $R^2(n)$ of the reconstructed displacements $\vec{u}^{(n)}(t)$ is shown for increasing number of modes used in the reconstruction.  This representation separates out the {\bf b}, $x$-component and {\bf c}, $y$-component of the displacement vector fields.}
\label{fig:recvec}
\end{figure*}

We can use the coefficients $c_m$ calculated with linear mode decomposition to reconstruct the crowd's dynamics.  If we only use a limited number $n$ of modes to define $\vec{u}^{(n)}(t) = \sum_{m = 1}^{n} c_m \vec{e}_m$ as a partial reconstruction of the dynamics, we can better understand the different contributions coming from low- and high-$m$ modes.  In the main text (Fig.~2), we were primarily concerned with (i) the role of the first mode, (ii) low-$m$ modes (modes up to $m = 30$), and (iii) convergent modes (modes up to $m = 190$).  Here, we go further and show reconstructions of the crowd's dynamics that incorporates non-convergent modes (modes up to $n = 300$) and a complete reconstruction using all 556 modes (Supplementary Fig.~\ref{fig:recvec}a).  Together with results in the main text, we have a more complete, yet still-qualitative, picture for how the projection coefficients contribute to the reconstructions.

To quantify how much of the reconstructed dynamics $\vec{u}^{(n)}(t)$ reproduces the actual crowd's displacements $\vec{u}(t)$, we use the common $R$-squared statistic, defined here as $R^2(n) =  1 - SS_{n} / SS_{{\rm Total}}$, where $SS_{n} = \sum_{i = 1}^N ( \vec{u}_i - \vec{u}_i^{(n)})^2$, $SS_{{\rm Total}} = \sum_{i = 1}^N ( \vec{u}_i - \langle \vec{u} \rangle)^2$, and $i$ runs over all grid-points.  While the formal notation drops the component index and time-dependence for convenience, the actual calculation is done by separating the $x$- and $y$-components (Supplementary Fig.~\ref{fig:recvec}b and c).  As anticipated, we see $R^2(n = 1)$ peaks at the appearance of the density wave at $t = 190$ and at frame index $t \approx 50$ and 150 (Supplementary Fig.~\ref{fig:recvec}b and c, blue lines), confirming the substantial role played by the first mode during moments of coherent collective motion.  Looking more broadly at $R^2$ throughout $T$, we see that low-$m$ modes are generally insufficient for fully reconstructing the dynamics (Supplementary Fig.~\ref{fig:recvec}b and c, blue and orange lines during $T$).  This finding is not unexpected because low-$m$ modes convey information about coherent collective motion, whereas turbulent crowd dynamics features a degree of spatial randomness.  Because $M$ convergent modes perfectly reconstruct the dynamics of each of the $M$ frames used to calculate them, we find the $n = 190$ modes fully reconstruct the dynamics during $T$ (Supplementary Fig.~\ref{fig:recvec}b and c, yellow lines).  Conversely, $R^2$ drops significantly at the beginning of $W$ when $t > 190$, and all modes are needed in order to reconstruct $\vec{u}(t)$ (Supplementary Fig.~\ref{fig:recvec}b and c, green lines).

\subsection{Behavior of the squared projection coefficients}

\begin{figure*}
\centering
\includegraphics{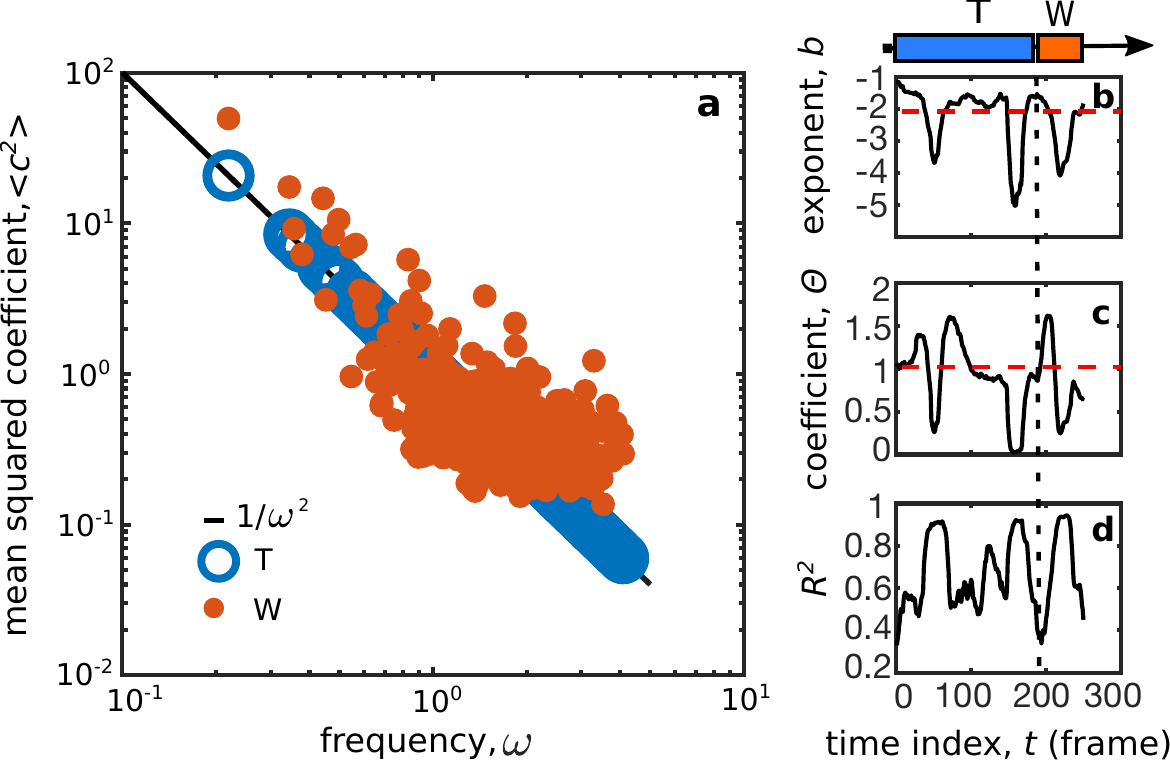}
\caption{\small Statistical properties of the projection coefficients.  {\bf a} The mean squared coefficients $\langle c_m^2 \rangle$ averaged over the time frames constituting $T$ (blue circles) and $W$ (orange dots) plotted against each modes' frequency. Time series of {\bf b} exponent, {\bf c} coefficient, and {\bf d} $R^2$ of the power law fit to $c_m^2$ averaged over a moving window. The red dashed horizontal lines in {\bf b} and {\bf c} are theoretically predicted values $\Theta_T = 1$ and $b_T = -2$.}
\label{fig:c2}
\end{figure*}

Having extracted the coefficients $c_m(t)$ with linear mode projections, we now seek an interpretive framework for analyzing their time-averaged and time-dependent behavior.  A useful model system comes from analyzing the statistical properties of an equilibrium non-interacting classical system of phonons.  In this minimal toy model, the eigenvalues $\lambda_m$ are related to the mode's vibrational frequency according to $\omega_m = 1/\sqrt{\lambda_m}$, and $c_m^2$ are the mode's occupation numbers.  To develop an expression for the distribution of $c_m^2$, we write a 1D Fourier representation of the system as $\Psi = \sum_{m=1}^{2N} c_m \sin(k_m x - \omega_m t)$, where $k_m$ is the wave number of mode $m$.  Here, the time-averaged energy is $\langle E \rangle \propto \langle \dot{\Psi}^2 \rangle \propto \sum_{m=1}^{2N} c_m^2 \omega_m^2$, which is independent of time because equipartition holds for the case being considered.  The relation between coefficients and frequencies is therefore $c_m^2 = \Theta/\omega^2_m = \Theta \lambda_m$, where $\Theta$ is the effective temperature.  Though derived for a much simpler system, this expression is useful for interpreting the distribution and behavior of the projection coefficients.

Interestingly, the same power law relation is expected to be a time-average property of any system when the time frames used to compute the eigenmodes are the \textit{same as those} used to time-average the squared coefficients of the projection on these eigenmodes.  In our case, the displacements $\vec{u}(t)$ are measured on a fixed grid, allowing the covariance matrix to be written in the compact form $C_{ij} = \langle \vec{u}_i(t) \cdot \vec{u}_j(t) \rangle_T$, where the average is computed using the set of time frames during turbulent motion $T$.  $C_{ij}$ is real, positive, and thus can be diagonalized as $C = \vec{e} \ \Lambda \ \vec{e}^{\,T}$, where $\vec{e}$ are the modes of the system and $\Lambda$ is the diagonal eigenvalue matrix. Reducing notational complexity by suppressing explicit $x,y$ coordinates in the superscripts, we find
\begin{eqnarray}
\big\langle c_m(t) c_n(t)  \big\rangle_T &=& \sum_{ij}  \big\langle u_i(t) e^i_m  u_j(t) e^j_n   \big\rangle_T, \nonumber \\
  &=& \sum_{ij}  \big\langle u_i(t)   u_j(t)  \big\rangle_T   e^i_m e^j_n, \nonumber \\
  &=& \sum_{ij}  C_{ij}   e^i_m e^j_n, \nonumber \\
  &=& \sum_{ij}  e^q_i \Lambda_{qk} e^k_j    e^i_m e^j_n, \nonumber \\
  &=& \delta^q_{m} \delta^k_{n} \Lambda_{qk}, \nonumber \\
  &=& \Lambda_{mn}, \nonumber 
\end{eqnarray}
where $i$ and $j$ index grid points, and $\delta^q_{m}$ is the Kroneker delta.  Because $\Lambda$ is diagonal, this calculation implies the power law relation $\langle c_m^2(t) \rangle_T = \lambda_m = 1/\omega_m^2$ holds by construction.  Conversely, when $c_m^2(t)$ is computed by averaging over a subset of time frames smaller than $T$, a measurement that reproduces this power law relationship would signal equilibrium or equilibrium-like behavior of the system during the time frames sampled.  

Turning now to our data, we averaged $c_m^2$ over all frames during $T$ and plotted this quantity against $\omega_m$ (Supplementary Fig.~\ref{fig:c2}a, blue circles).  As expected, we found the data fit a power law of the form $\langle c^2 \rangle_T = \Theta_T \omega^{b_T}$, where $\Theta_T = (1.0 \pm 0.0)$, $b_T = (-2.0 \pm 0.0)$, and the mode subscript index has been suppressed for clarity (Supplementary Fig.~\ref{fig:c2}a, black line, $R^2 = 1.00$).  

Next, we computed and averaged $c_m^2(t)$ over the time frames of $W$ and found the distribution $\langle c_m^2(t) \rangle_W$ deviates slightly from power law behavior, indicating an increased relevance of both low- and high-$m$ modes (Supplementary Fig.~\ref{fig:c2}a, orange dots).  

We then boxcar averaged $c_m^2(t)$ over a sliding window of 20 frames and fitted this moving average of the distribution to the same power law.  This calculation produced the time-dependent fit parameters $\Theta(t)$ and $b(t)$, offering insights on \textit{if}, \textit{when}, and \textit{how} the system deviates from equilibrium-like behavior (Supplementary Fig.~\ref{fig:c2}b and c).  We observed similar temporal patterns in both $\Theta(t)$ and $b(t)$: they oscillate around $\Theta_T = 1$ and $b_T = -2$ and exhibit sudden fluctuations around $t \approx 50,$ 150, and after 190 corresponding to the power injections discussed in the main text as well as the density wave (Fig.~3b).  We also find these fluctuations correspond to higher values of $R^2$ in the power law distribution fit (Supplementary Fig.~\ref{fig:c2}d).  This observation is likely due to the fact that power law fits of noisy data are dominated by data scatter at the extreme ends of the fitting domain.  In our case, we know contributions from the $m = 1$ mode are significant during power injections and the wave, so it is reasonable to attribute the increase in $R^2$ not to a better overall fit, but to a more skewed distribution.

\subsection{Additional results on the explanatory power $\alpha_m^2$} 

\begin{figure*}
\centering
\includegraphics{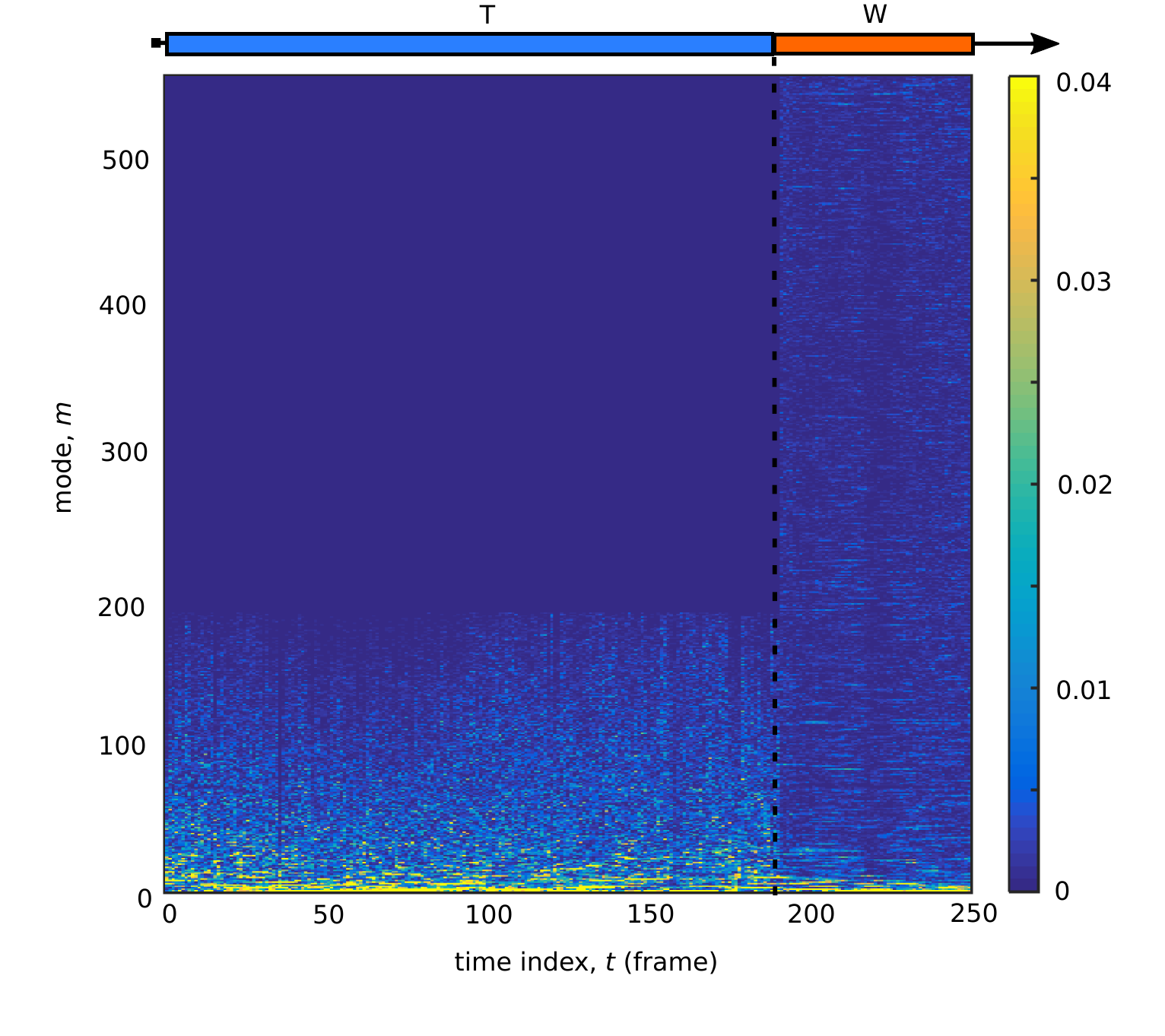}
\caption{\small Time series of the explanatory power $\alpha^2_m(t)$ up to $m = 556$.  The scale of the heat-map coloring was adjusted 10-fold relative to the main text in order to better visualize the smaller contributions from high-$m$ modes after $t = 190$.}
\label{fig:a2all}
\end{figure*}

It is instructive to look at the time series of $\alpha^2_m(t)$ for all values of $m$ up to 556 (Supplementary Fig.~\ref{fig:a2all}).  In this plot, we see two distinct boundaries: one dividing time index $t$ at $t = 190$, and the other dividing mode number $m$ at $m = 190$ for $t \le 190$.  As hinted in our discussion about reconstructing the crowd's dynamics (Supplementary Fig.~\ref{fig:recvec}b, c) and in the analytical derivation of $\big\langle c^2_m(t) \big\rangle_T$, the quantitative matching of these boundaries is due to the fact that we use 190 time frames to compute the correlation matrix, which gives 190 convergent modes.  Thus, for $t \le 190$ the remaining 366 modes are, by construction, negligible for reconstructing the dynamics (Supplementary Fig.~\ref{fig:a2all}, dark blue area for $m > 190$; Supplementary Fig.~\ref{fig:recvec}b and c).  If we had instead used 150 frames to compute the modes, then $m$ would have a boundary at $m = 150$ (see discussion below in ``Stability of the analysis with respect to the length of video data'').  Therefore this boundary on mode number is related to the amount of frames used to compute the modes, and not specifically to the kind of collective motion observed in the crowd.

Conversely, after $t = 190$, the explanatory power across high-$m$ modes is more uniformly distributed (Supplementary Fig.~\ref{fig:a2all}, especially $m > 190$).  This change in the distribution of high-$m$ modes is not a consequence of the transition from turbulent motion to the collective density wave, but instead, a consequence of the fact that we stop mode analysis at $t = 190$.  Therefore, displacements occurring at $t > 190$ correspond to time frames that are in the ``future'' with respect to the modes.  Fortunately, we find low-$m$ modes are relevant for describing these ``future displacements'' as demonstrated by the power concentration in low-$m$ modes reflecting the presence of the density wave at $t > 190$.  While for granular media the relevance of modes to describe future displacements is an established fact~\cite{chen2011measurement, manning2011vibrational}, it was not guaranteed to hold for dense human crowds.  Overall, the fact that low-$m$ modes retain their significance in the ``future'' makes them useful for forecasting collective motion with quantitative accuracy.

\subsubsection{Temporal auto-correlation} 

\begin{figure*}
\centering
\includegraphics{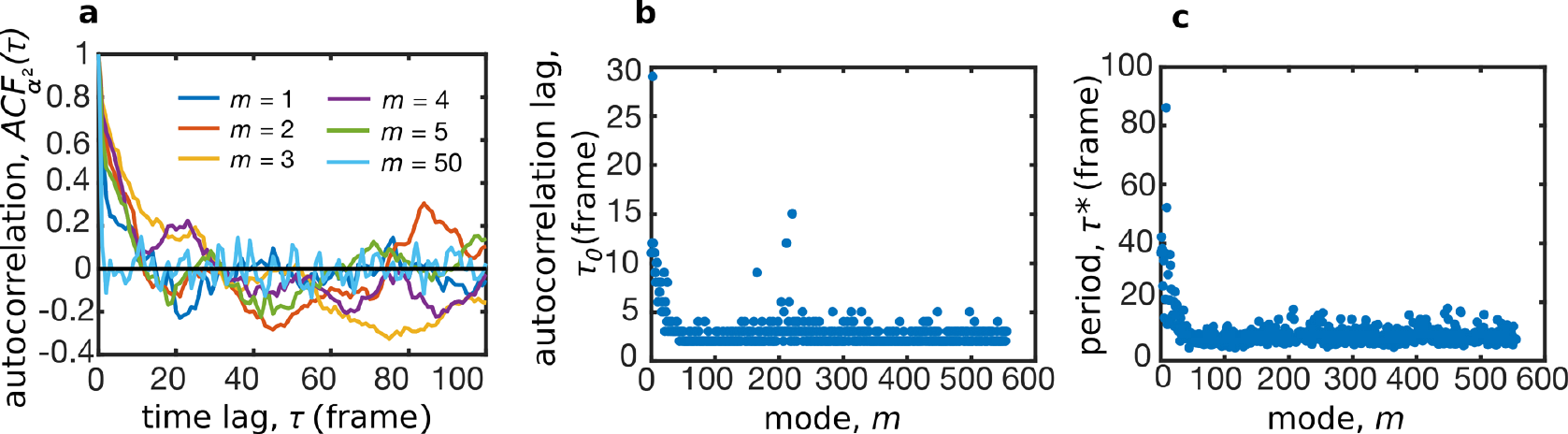}
\caption{\small {\bf a}, Autocorrelation function of the explanatory power $\alpha_m^2(t)$ as a function of time lag for selected low-$m$ and high-$m$ modes. {\bf b}, Autocorrelation lag and {\bf c}, oscillation period as a function of mode number.}
\label{fig:autocorr}
\end{figure*}

We also studied the behavior of the explanatory power through the standard autocorrelation function $ACF_{\alpha^2}(\tau) = \langle \alpha^2_m(t) \alpha^2_m(t+\tau) \rangle$, where $\tau$ is the time lag.  In signal processing theory, the autocorrelation function is often used to detect a signal's periodicity and decay rate in presence of strong noise~\cite{box2015time}. Here, we looked at the autocorrelation decay lag $\tau_0$, defined as the smallest time lag at which the autocorrelation is equal to zero, and at the period $\tau^*$ of the $ACF_{\alpha^2}$ during turbulent crowd motion.  The autocorrelation function showed oscillatory behavior for low-$m$ modes, while for high-$m$ modes it was peaked only at $\tau = 0$ and assumed near-zero values otherwise (Supplementary Fig.~\ref{fig:autocorr}a).  We found the autocorrelation time $\tau_0$ and oscillation period $\tau^*$ decrease with mode number (Supplementary Fig.~\ref{fig:autocorr}b and c), consistent with the increasing spatial randomness of higher-$m$ modes (Supplementary Fig.~\ref{fig:spatial}b).  In particular, $\tau_0 = 11$ and $\tau^* = 36$ frames for $\alpha^2_1(t)$, corresponding to $\approx 1.3$ and $4.5$ seconds respectively.  Periodicity and long autocorrelation in $\alpha^2_m(t)$ for low-$m$ modes strongly suggest the explanatory power metric is an excellent candidate for forecasting collective behavior over longer periods of time.

\section{Candidate forecasting measures}

A general remark concerning the analyzed concert footage is that during turbulent motion we observed two minor fluctuations along $\vec{e}_1$, at $t \approx 50$ and 150, that decayed within a few seconds, and one large-scale fluctuation, the density wave, that amplified and propagated across the audience.  Fundamental to forecasting the emergence of collective motion is to understand why some fluctuations dissipate while others amplify and propagate.  In this section, we propose two measures based on the $\alpha^2_m(t)$ time series that shed light the system's dynamics and lead us toward a clearer understanding of the patterns that forecast collective motion.

\begin{figure*}
\centering
\includegraphics{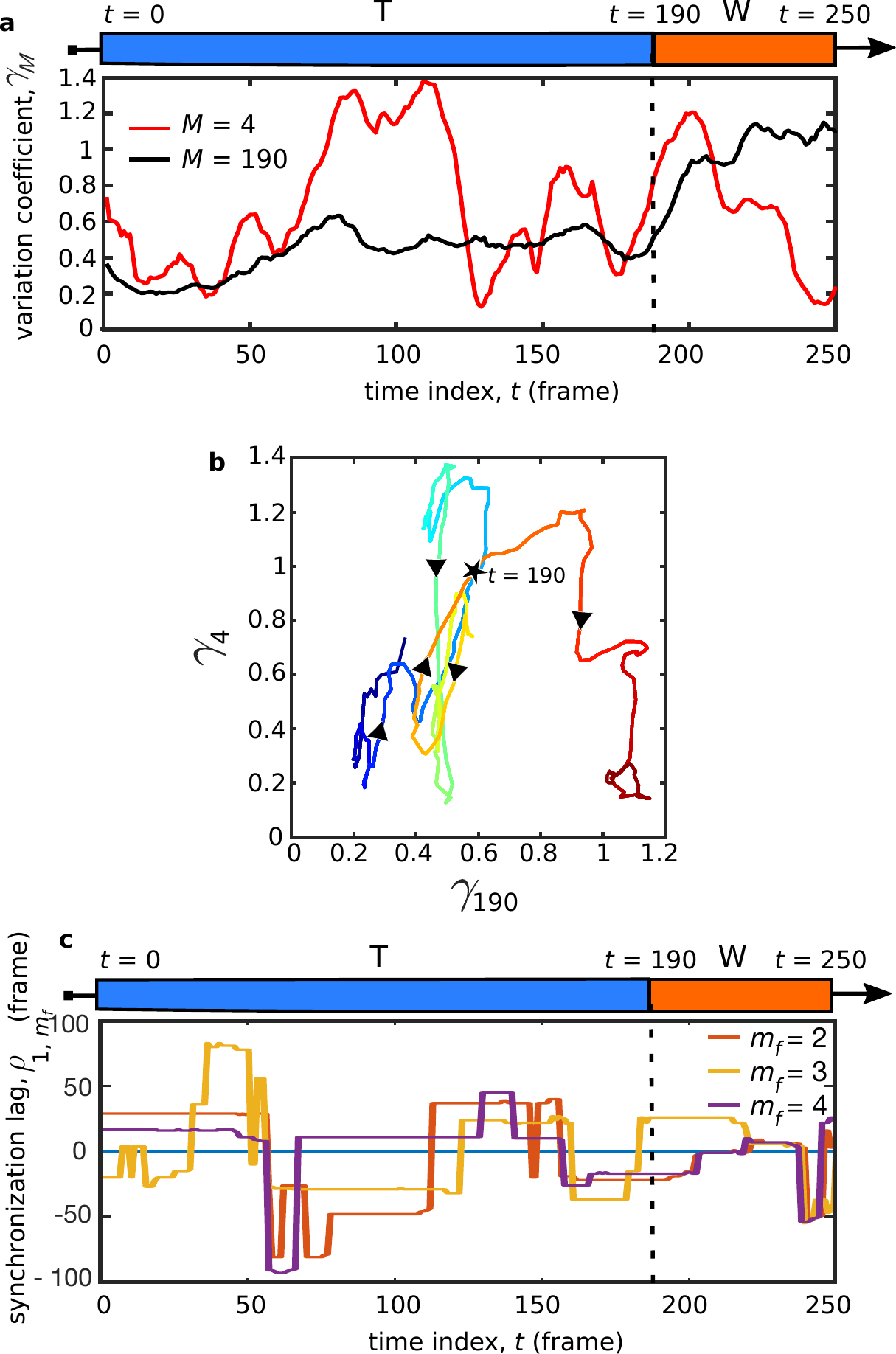}
\caption{\small Candidate measures for forecasting collective motion. {\bf a}, Variation coefficients $\gamma_4(t)$ and $\gamma_{190}(t)$. {\bf b}, Visualization of the system's dynamics in the phase space of the variation coefficients $\langle \gamma_{190}$, $\gamma_4 \rangle$.  Time flows from blue to red, and the black star separates $T$ from $W$. {\bf c}, Identification of synchronous modes: time evolution of the synchronization lag (within $\pm 100$ time frames) between $m_i = 1$ and $m_f = 2, 3,$ and 4.}
\label{fig:gamma}
\end{figure*}

\subsection{The variation coefficient}

The first measure for forecasting collective motion of dense crowds aims at detecting (i) perturbations in the distribution of $\alpha^2_m(t)$ and (ii) power injections and cascades in the low-$m$ modes associated with the appearance of the density wave.  First, we smooth the time series of $\alpha^2_m(t)$ by boxcar averaging over 20 time frames.  The ratio $\Upsilon(t, m) = \alpha^2_m(t)/ \langle \alpha^2_m(t)\rangle_{t<50}$ compares the behavior of the time series in the numerator with a reference state in the denominator, which in this case was the average explanatory power prior to the first energy injection at $t \approx 50$.  The standard deviation $\sigma_{M}(t) = {\rm STD}\big[ \Upsilon(t, m)\vert_{m < M} \big]$ quantifies heterogeneity of the first $M$ modes' deviations from the reference state.  In particular a large standard deviation indicates that some modes are deviating significantly more than others, which is what happens during power injections and wave propagation.  To compare this heterogeneity over different ranges of $M$ with each other, we normalize $\sigma_M$ and define the variation coefficient $\gamma_{M}(t) = \sigma_{M} / \mu_{M}$, where $\mu_{M}(t) = \langle \Upsilon(t, m) \rangle_{m < M}$ is the mean deviation from the reference state of the first $M$ modes.  

We used the variation coefficient $\gamma_M(t)$ to study data from the \textit{Oasis} concert. 
We monitor the behavior of the first four modes during power injections and cascades by using $\gamma_4(t)$, and we compare it to the behavior of the convergent modes by using $\gamma_{190}(t)$ (Supplementary Fig.~\ref{fig:gamma}a).  We find $\gamma_4(t)$ peaks weakly at $t \approx 50$, 150, and strongly at the appearance of the density wave. It also shows a prolonged peak between the two power injections at $t \approx 50$ and 150, indicating sensitivity to power cascades. While $\gamma_{190}(t)$ does not distinguish single energy injections or cascades, it shows a gradual increase over time, signaling increasing heterogeneity in the deviations from the average explanatory power distribution for $t<50$. The sudden increase at the transition between $T$ and $W$ reflects the fact that power is more broadly distributed among higher-$m$ modes (Supplementary Fig.~\ref{fig:a2all}). $\gamma_{4}(t)$ also peaks at the transition between $T$ and $W$ due to the power cascade into $m = 1$  (Fig.~3b).  Finally, for $t > 210$, we observe a drop in $\gamma_{4}(t)$ signaling diminished power exchange between modes which is mainly concentrated in mode $m = 1$. 

An alternative for visualizing the evolution of the system is to plot its trajectory in the $\langle \gamma_{190}$, $\gamma_{4} \rangle$ phase space (Supplementary Fig.~\ref{fig:gamma}b).  Here, crowd turbulence is represented by an elongated ellipse, while the propagation of the density wave causes a substantial deviation off this trajectory.  Taking inspiration from dynamical systems, one can conceive of this trajectory as being driven by an underlying phase portrait that evolves in time. 

Overall, if trained on further data sets, the variation coefficient $\gamma_M(t)$ combined with the total linear power $A^2(t)$ could constitute the core of an algorithm for real-time crowd monitoring and early warning signal detection of potentially dangerous collective motion.

\subsection{Detecting synchronous modes}

The second measure we consider for forecasting collective motion of dense crowds explores cross-correlations among the $\alpha_m^2(t)$ time series.  For each mode $m_i$ and any other mode $m_f$ we fix a time $t^*$ and compute the cross-correlation between $\alpha^2_{m_i}(t^*)$ and the whole time series of $\alpha^2_{m_f}(t)$ within a symmetric window of $\pm 100$ time frames.  We define the ``synchronization lag'' $\rho_{m_i, m_f}(t^*)$ as the time-lag corresponding to the maximum of the cross-correlation function over the window of $\pm 100$ frames.  A synchronization between $m_i$ and $m_f$ implies $\rho_{m_i, m_f}(t) = 0$, and thus the time series of $\rho_{m_i, m_f}(t)$ can be used to identify transient synchronizations.  This metric can also be used to retrospectively analyze the evolution of synchronization: $\rho_{m_i, m_f}(t)$ approaching zero means modes $m_i$ and $m_f$ are becoming more coordinated.  

As an example, we computed $\rho_{m_i, m_f}(t)$ between the first mode $m_i = 1$, where we observe power injections, and modes $m_f = 2, 3,$ and 4, which are the modes involved in power cascades (Supplementary Fig.~\ref{fig:gamma}c and Fig.~3b).  We notice that synchronization lags oscillate between positive and negative values.  Times of maximal synchronization between modes from $m = 2$ to 4 seem to correlate with power injections.  The amplitude of oscillations decreases in time, meaning that the analyzed modes become more coordinated with $m = 1$, that is, their power exhibits increasingly similar behavior over time.  It is also interesting to observe that the different synchronization lags assume negative and positive values at approximately the same time, and seem to progressively converge toward a common trend. This is particularly evident for modes 2 and 4, that become fully coordinated around $t = 190$.  Although the currently available data do not allow us to draw conclusions on which pattern might be relevant to forecast density waves, it is clear that the synchronization lag identifies underlying patterns in the $\alpha^2_m(t)$ time-series.

\begin{figure*}
\centering
\includegraphics{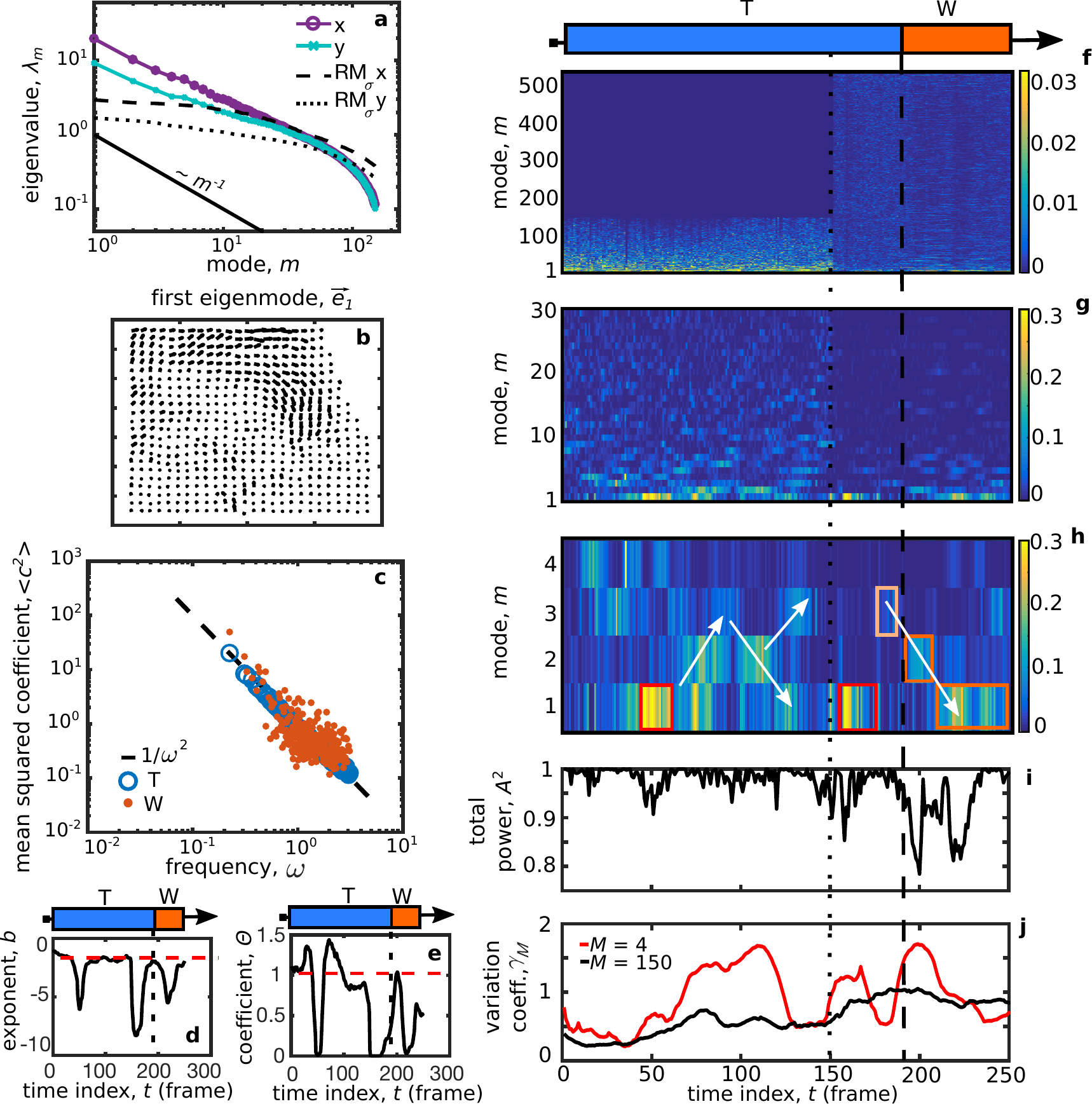}
\caption{\small Main results are reproduced when modes are calculated by using only the first 150 time frames. Here we examine the {\bf a}, Eigenvalue spectrum, {\bf b}, first mode $\vec{e}_1$, {\bf c}, average squared projection coefficients, {\bf d}, fitted power law exponent, {\bf e}, fitted power law coefficient, and see all of these measures are consistent with results calculated using all 190 time frames.  We also examine the explanatory power for {\bf f}, all 556 modes, {\bf g}, the first 30 modes, {\bf h}, and the first 4 modes to discover what effect the choice of how many times frames to analyze has on the robustness of our detection of power injections and cascades.  {\bf i} The total linear power {\bf j}, and variation coefficients are essentially unchanged from our earlier findings.}
\label{fig:time150}
\end{figure*}

\section{Stability of the analysis with respect to the length of video data}

We repeated our analysis with a shorter segment of video data that ends 40 frames \textit{before} $W$ starts.  By restricting the computation of the covariance matrix $C_{ij}$ to the first 150 frames of $T$, we can test whether our detection of the power cascade at $t = 190$ is an artifact arising from the choice of frames included in our analysis (Supplementary Fig.~\ref{fig:time150}). The main results show that our analysis is robust for changes in the length of the time series of video data.  In particular, we observe that the eigenvalue spectrum trend is unchanged for the lowest energy modes, and the $m = 1$ mode is still a wave directed towards and along the security barrier (Supplementary Fig.~\ref{fig:time150}a and b).  The distribution of squared projection coefficients retains a power law behavior when averaged over the full time interval used to compute the mode basis.  Similarly, fitting the distribution on a moving window shows deviations from this power law appear during minor collective fluctuations at $t \approx 50,$ 150, and during wave propagation (Supplementary Fig.~\ref{fig:time150}c-e).  Power cascades in $\alpha_m^2(t)$ are present as in the original analysis (Supplementary Fig.~\ref{fig:time150}f-h), and as expected, the 150 time frames used for computing $C_{ij}$ result in 150 convergent modes sufficient for describing motion during $t \le 150$.  For $t > 150$, we again find that the explanatory power is broadly distributed among all modes.  The total linear power $A^2(t)$ (Supplementary Fig.~\ref{fig:time150}i) is almost unchanged with respect to the analysis presented in the main text.  Finally, the variation coefficients $\gamma_4(t)$ and $\gamma_{190}(t)$ show similar trends to the previous analysis, though more refinement of these metrics is still generally needed.  Summarizing these results, we conclude the essential findings presented in the main text do not depend on our choice of including the first 190 frames, and the cascades and signatures of non-linearity that appear to forecast collective motion are robust signals worth further study.

%\vfill \eject
%\bibliography{DynamicModes}
%\bibliographystyle{unsrt}

\vfill \eject

\end{document}